\documentclass[twocolumn,letterpaper,amsmath,amssymb,floatfix,aps,superscriptaddress]{revtex4}

\usepackage{graphicx}
\usepackage{dcolumn}
\usepackage{bm}
\usepackage[usenames]{color}
\usepackage{hyperref}
\usepackage{ulem}

\begin{document}

\title{Driven translocation of a semi-flexible polymer through a nanopore}

\author{Jalal Sarabadani}
\email{jalal.sarabadani@aalto.fi}
\affiliation{Department of Applied Physics and COMP Center of Excellence, Aalto University School of Science, 
P.O. Box 11000, FI-00076 Aalto, Espoo, Finland}

\author{Timo Ikonen}
\affiliation{~VTT Technical Research Centre of Finland Ltd., P.O. Box 1000, FI-02044 VTT, Finland}

\author{Harri M\"okk\"onen}
\affiliation{Department of Applied Physics and COMP Center of Excellence, Aalto University School of Science, 
P.O. Box 11000, FI-00076 Aalto, Espoo, Finland}

\author{Tapio Ala-Nissila}
\affiliation{Department of Applied Physics and COMP Center of Excellence, Aalto University School of Science, 
P.O. Box 11000, FI-00076 Aalto, Espoo, Finland}
\affiliation{Department of Physics, Box 1843, Brown University, Providence, Rhode Island 02912-1843.}

\author{Spencer Carson}
\affiliation{Department of Physics, Northeastern University, Boston MA 02115}

\author{Meni Wanunu}
\affiliation{Department of Physics, Northeastern University, Boston MA 02115}

\date{December 27, 2016}

\begin{abstract}
We study the driven translocation of a semi-flexible polymer through a nanopore by means of a modified version of
the iso-flux tension propagation theory (IFTP), and extensive molecular dynamics (MD) simulations. We show that
in contrast to fully flexible 
chains, for semi-flexible polymers with a finite persistence length $\tilde{\ell}_p$ the
 {\it trans} side friction must be explicitly taken into account to properly
describe the translocation process. In addition, the scaling of the end-to-end distance $R_N$ 
as a function of the chain length $N$ must be known. 
To this end, we first derive a semi-analytic scaling form for $R_N$, which reproduces the limits of
a rod, an ideal chain, and an excluded volume chain in the appropriate limits. 
We then quantitatively characterize the nature of the {\it trans} side friction based
on MD simulations of semi-flexible chains. Augmented with these two factors, 
the modified IFTP theory shows that there are three main regimes for the scaling of the average 
translocation time $\tau \propto N^{\alpha}$. In the stiff chain (rod) limit $N/\tilde{\ell}_p \ll 1$, 
{$\alpha = 2$}, which
continuously crosses over in the regime $ 1 < N/\tilde{\ell}_p < 4$ towards the ideal chain behavior with
{$\alpha = 3/2$}, which is reached in the regime $N/\tilde{\ell}_p \sim 10^2$. Finally, 
in the limit $N/\tilde{\ell}_p \gg 10^6$ 
the translocation exponent approaches its asymptotic
value $1+\nu$, where $\nu$ is the Flory exponent.  Our results are in good agreement with
available simulations and experimental data.
\end{abstract}

\maketitle

{\it Introduction} -- Since the seminal works by Bezrukov {\it et al}. \cite {Parsegian} in 1994, 
and two years later by Kasianowicz {\it et al}. \cite{Kasianowicz}, polymer translocation through 
nanopores has become one of the most active 
research topics in soft condensed matter physics \cite{Muthukumar_book,Milchev_JPCM,Tapio_Review}. 
It plays an important role in many biological processes such as virus injection and protein transportation through membrane channels 
\cite{Alberts}. It also has many technological applications such as drug delivery \cite{Meller_JPhysCondMatt}, gene therapy and rapid
DNA sequencing \cite{Kasianowicz,Meller_PRL_2001,Aksimentiev_NanoLett_2008,Tapio_PRL_2008,Golestanian_PRX_2012}, and has 
been motivation for many experimental and theoretical studies \cite{Tapio_Review,storm2005,branton2008,Spencer2014,Stein2014,sung1996,muthu1999,%
metzler2003,kantor2004,grosberg2006,sakaue2007,luo2008,luo2009,schadt2010,Milchev_JPCM,rowghanian2011,Muthukumar_book,bhatta2009,%
dubbeldam2012,sakaue2010,saito2012a,ikonen2012a,ikonen2012b,jalal2014,jalal2015,ikonen2013,slater2008a,slater2008b,Lam_JStatPhys2015,dubbeldam2014,%
AniketJCP2013,AniketPolymerSciSerC2013,ClementiPRL2006,DaiPolymers2016,NetzEPL2006}.

\begin{figure}[b]
\begin{center}
        \includegraphics[width=0.21\textwidth]{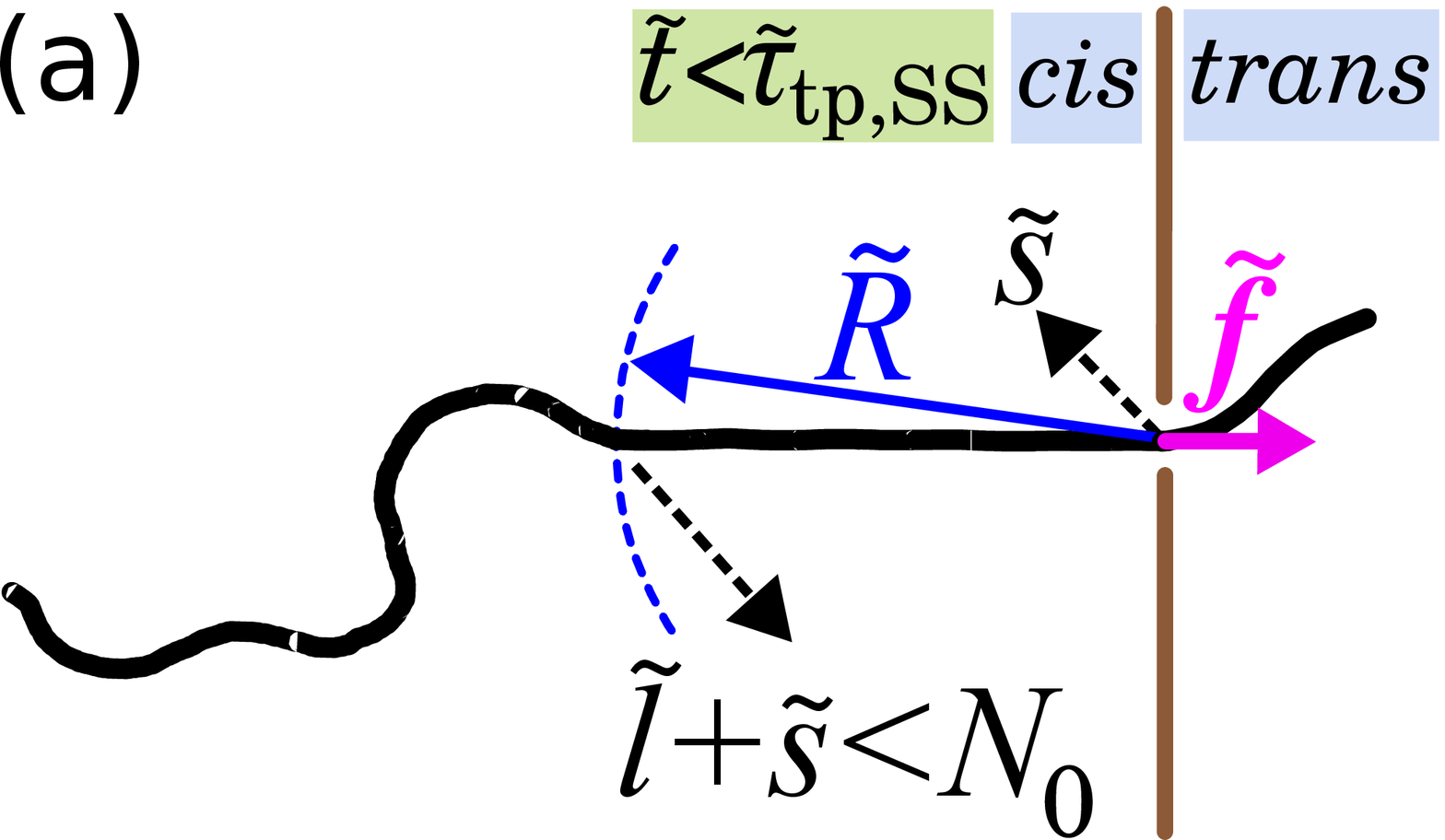}
    \hskip0.4cm
        \includegraphics[width=0.227\textwidth]{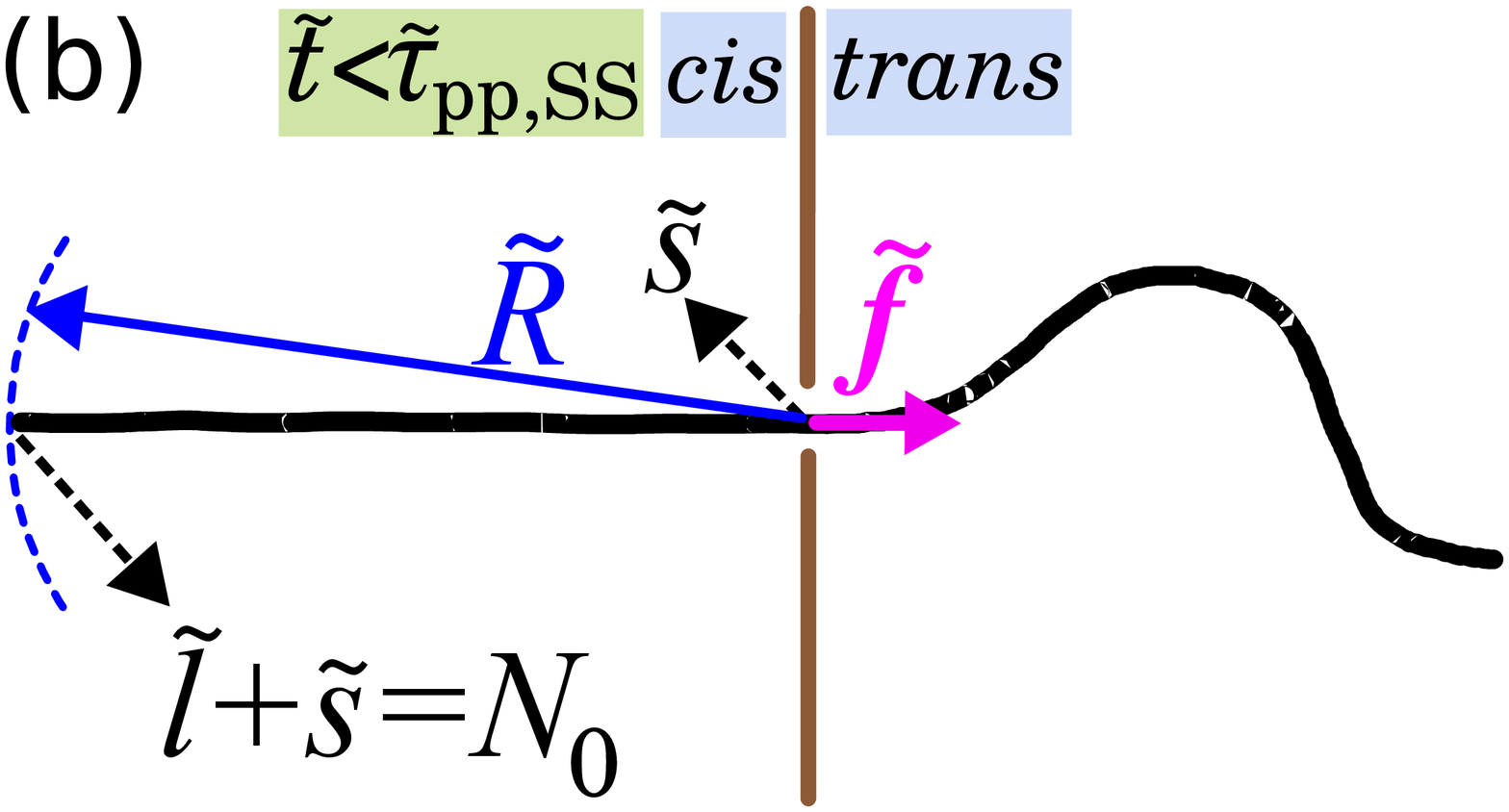}
\caption{
(a) A schematic of the translocation process in the tension propagation (TP) stage, 
i.e. $\tilde{t} < \tilde{t}_{\textrm{tp,SS}}$, 
for the strong stretching (SS) regime. $N_0$ is the contour length of polymer 
and the translocation coordinate $\tilde{s}$ equals 
the number of beads that have already been translocated into the {\it trans} side. The number of beads influenced by the 
tension force is $\tilde{l} + \tilde{s}$, which during TP 
stage is less than $N_0$. $\tilde{R}$ determines the location of the tension front. 
(b) The translocation process for SS  
regime during the post propagation stage where the tension front has reached the chain end, 
which yields $\tilde{l} + \tilde{s}= N_0$.
} 
\label{fig:schimatic_1}
\end{center}
\end{figure}

Most analytical and theoretical studies to date have focused on field-driven translocation of flexible polymers
through nanopores. A break-through in this problem came from
Sakaue, who employed the idea of
{\it tension propagation} (TP) to explain the physical mechanism of the driven translocation process 
\cite{sakaue2007}. According to TP theory when the external driving force, which is due to an external electric field
across the pore, acts on the bead(s) at the pore in the direction of 
{\it cis} to {\it trans} side (see Fig.~\!\ref{fig:schimatic_1}), a tension front propagates along 
the backbone of the chain in the {\it cis} side of the chain. Consequently, the {\it cis} side is divided 
into mobile and immobile parts, where the mobile part of the chain has been already influenced by the tension 
force and moves towards the pore, and the immobile part of the chain is in its equilibrium state, i.e. its average velocity 
is zero. 

Following Sakaue's work, in a series of papers Ikonen {\it et al.} developed a
Brownian dynamics - TP theory (BDTP) to take into account the effect of pore friction, finite chain length, and
thermal fluctuations due to the solvent during the course of translocation \cite{ikonen2012a,ikonen2012b}. 
Most recently, the BDTP theory was reformulated within the constant monomer {\it iso-flux} approximation \cite{rowghanian2011} (IFTP)  
\cite{jalal2014,jalal2015}, leading to a fully quantitative and self-consistent theory of dynamics of
driven translocation with only one free parameter, the effective pore friction.
A key role in the theory is played by the total effective friction, which comprises the constant pore friction (interaction of the monomers
with the nanopore) and drag from the {\it cis} part of the chain. For fully flexible chains, the contribution from the {\it trans} side of 
the friction can be included in the pore friction, and need not be explicitly considered.

However, in many cases of practical interest the translocating polymers are not fully flexible -- e.g. for double-stranded DNA, the
persistence length $\ell_p$ is typically about 500 {\AA}. To unravel the influence of stiffness to translocation, 
in this Letter we consider the pore-driven translocation dynamics of 
semi-flexible polymers 
with a finite persistence length within the IFTP theory. 
We argue that unlike for fully flexible chains, the {\it trans} side 
friction has a significant contribution to the dynamics and must be explicitly added
to the expression for the total friction. To calculate the chain drag, 
we derive a semi-analytic form for the end-to-end scaling form $R_N$ for
semi-flexible chains, which correctly incorporates the various scaling regimes and crossover
between them for different ratios
of the persistence and chain lengths $\tilde{\ell}_p/N$.
Neither of these factors have been considered in the previous works 
\cite{AniketJCP2013,AniketPolymerSciSerC2013,ClementiPRL2006,DaiPolymers2016}.
When properly augmented with the correct end-to-end scaling form and time-dependent {\it trans} side friction,
the IFTP theory shows that the average translocation time displays complex scaling and crossover behavior as
a function of $\ell_p/N$. In the appropriate limits, the IFTP
theory also recovers the exactly known results for the scaling exponent of the translocation time.
It is important to note that in the IFTP theory there is only one unknown parameter,
the {\it effective pore friction} $\eta_{\textrm{p}}$, which can be obtained either
experimentally or from MD simulations \cite{ikonen2012a,ikonen2012b,jalal2014,jalal2015}.

{\it Theory: (a) Strong stretching regime} -- In the IFTP theory,
we use dimensionless units denoted by tilde 
as $\tilde{X} \equiv X / X_u$, with the units of length $s_u \equiv \sigma$, time $t_u \equiv \eta \sigma^2 / (k_B T)$, 
force $f_u \equiv k_B T/\sigma$, velocity $v_u \equiv \sigma/t_u = k_B T/(\eta \sigma)$, 
friction $\Gamma_u \equiv \eta$, and monomer flux $\phi_u \equiv k_B T/(\eta \sigma^2)$,
where $\sigma$ is the segment length, $T$ is the temperature of the system, $k_B$ is the Boltzmann constant, and 
$\eta$ is the solvent friction per monomer.
The quantities without the tilde, such as the force, friction and length,
are expressed in Lennard-Jones units.
In the overdamped Brownian limit \cite{ikonen2012a,ikonen2012b,jalal2014,jalal2015}, the
equation of motion for the translocation coordinate $\tilde{s}$ which is the number of beads 
in the {\it trans} side (see Fig.~\!\ref{fig:schimatic_1}), is given by 
\begin{equation}
\tilde{\Gamma} (\tilde{t}) \frac{d \tilde{s}}{ d \tilde{t}} =
\tilde{f} + \tilde{\zeta} (\tilde{t}) \equiv  \tilde{f}_{\textrm{tot}},
\label{BD_equation_1}
\end{equation}
where $\tilde{\Gamma} (\tilde{t})$ is the effective friction, and $\tilde{\zeta} (\tilde{t})$ is Gaussian white noise
which satisfies $\langle \zeta (t) \rangle = 0$ and $\langle \zeta (t) \zeta (t') \rangle = 2 \Gamma (t)
k_B T \delta (t - t ')$, $\tilde{f}$ is the external driving force, and $\tilde{f}_{\textrm{tot}}$ 
is the total force. 

In the iso-flux assumption the monomers flux, 
$\tilde{\phi}\equiv d\tilde{s}/d\tilde{t}$,
on the mobile domain in the {\it cis} side and also through the pore is constant in space, but evolves in time 
\cite{rowghanian2011,jalal2014}. The tension front, which is the boundary between 
the mobile and immobile domains, is located at distance $\tilde{x}=-\tilde{R}(\tilde{t})$ from the pore. 
The external driving force acts on the monomer(s) inside the pore located at $\tilde{x} = 0$ (see Fig.~\!\ref{fig:schimatic_1}(a)).

It has been shown
\cite{ikonen2012a,ikonen2012b,jalal2014,jalal2015,ikonen2013}
that for flexible polymers the friction can be written as 
$\tilde{\Gamma} (\tilde{t}) = \tilde{\eta}_{\rm cis} (\tilde{t}) + \tilde{\eta}_{\textrm{p}}$,
and the translocation dynamics is essentially controlled by the time-dependent friction $\tilde{\eta}_{\rm cis} (\tilde{t})$ 
on the {\it cis} side of the chain, whereas the {\it trans} side friction is negligible and can be absorbed into the constant 
pore friction $\tilde{\eta}_{\textrm{p}}$. In the case of semi-flexible chains this approximation is not justified. Within the 
IFTP theory, the friction due to the {\it trans} side of the chain $\tilde{\eta}_{\rm TS}$ can be taken into
account as follows. In the strong stretching (SS) regime of strong driving,
where the mobile part of the chain in the {\it cis} side is fully straightened (cf. 
Figs.~\!\ref{fig:schimatic_1}(a) and (b)), we can integrate the force balance equation 
over the mobile domain {\cite{jalal2014}} and the monomer flux becomes
\begin{equation}
\tilde{\phi} (\tilde{t}) = \frac{\tilde{f}_{\textrm{tot}} (\tilde{t})}
{ \tilde{R} (\tilde{t}) +\tilde{\eta}_{\textrm{p}} + \tilde{\eta}_{_\textrm{TS}} }.
\label{phi_equation}
\end{equation}
By combining Eqs.~\!\eqref{BD_equation_1} and \eqref{phi_equation}, 
the effective friction is obtained as
\begin{equation}
\tilde{\Gamma} (\tilde{t}) = \tilde{R}(\tilde{t}) + \tilde{\eta}_{\textrm{p}} + \tilde{\eta}_{_\textrm{TS}}.
\label{Gamma_equation}
\end{equation}
The time evolution of $\tilde{s}$ is determined by Eqs. \eqref{BD_equation_1}, \eqref{phi_equation} 
and \eqref{Gamma_equation}, but knowledge of the position of the tension front on the {\it cis} side
of the chain $\tilde{R}(\tilde{t})$ is still required to find the full solution.
We will derive the equation of motion for $\tilde{R}(\tilde{t})$ separately for the TP 
and {\it post propagation} (PP) stages. In the TP 
stage the tension has not be reached the chain end
as presented in Fig.~\ref{fig:schimatic_1}~\!(a), while in the PP 
stage the final monomer has been already influenced by the tension force (see Fig.~\ref{fig:schimatic_1}~\!(b)).

\begin{figure}[b]
\begin{center}
    \includegraphics[width=0.38\textwidth]{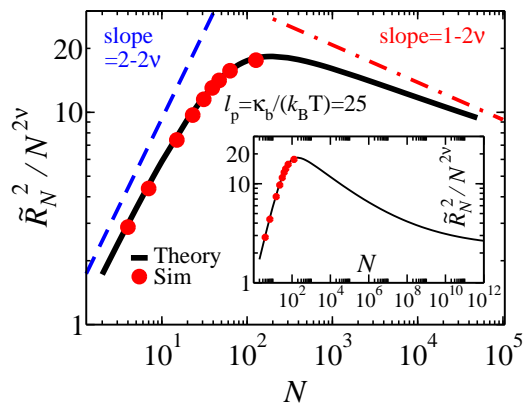}
\caption{
Normalized end-to-end distance $\tilde{R}_N^2/N^{2\nu}$ as a function of the contour length of the polymer $N$
for fixed value of bending rigidity (in the MD simulations) $\kappa_{{b}}= 30$, which 
corresponds to $\ell_p= 25$, when $k_B T=1.2$.
The black curve shows the analytical formula of Eq.~\!(\ref{end_to_end_distance})
while red dots present the MD simulations results. 
Inset shows crossover from Gaussian to self-avoiding behavior for an extended range of $N$.
} 
\label{fig:end_to_end_distance}
\end{center}
\end{figure}

{\it (b) End-to-end distance of a semi-flexible chain} --
To find the equation of motion for $\tilde{R} (\tilde{t})$, which is the root-mean-square of 
the end-to-end distance, an analytical form of {$\tilde{R} (\tilde{t})$ for} semi-flexible 
chains is needed. To this end, we have carried out extensive MD 
simulations of bead-spring models of semi-flexible chains in 3D. 
The technical details can be found in the Supplementary Material (SM).
The MD simulations have been done for different values of contour length $N \sigma$ and bending 
rigidity $\kappa_{{b}}$. In 3D the persistence length $\ell_{{p}}$ can be 
expressed as a function of $\kappa_{{b}}$ as $\ell_{{p}}= \kappa_{{b}}/(k_B T)$.
We find that the MD data (cf. Fig.~\!\ref{fig:end_to_end_distance}) is well described for 
all values of $N/\tilde{\ell}_{{p}}$ by the following semi-empirical analytic expression for the end-to-end distance 
of a semi-flexible chain:
\begin{eqnarray}
\tilde{R}_N = \bigg\{ && \hspace{-0.3cm} \tilde{R}^2_{F}
- \frac{\tilde{R}^4_{F} }{2 a_1 N^2 } \bigg[ 1- \exp \bigg( - \frac{2 a_1 N^2 }{ \tilde{R}^2_{F} }
\bigg) \bigg] \nonumber\\
&& \hspace{-0.5cm} + 2 \tilde{\ell}_{{p}} N \!
- \frac{2 \tilde{\ell}_{{p}}^{2}  }{b_1}  
\bigg[ 1 \!-\! \exp \bigg( \!\!- \frac{b_1 N }{ \tilde{\ell}_{{p}}} \bigg) \! \bigg] \! \bigg\}^{\frac{1}{2}}.
\label{end_to_end_distance}
\end{eqnarray}
Here {$\tilde{R}_{F}= A \tilde{\ell}_{{p}}^{\nu_{{p}}} N^{\nu}$,}  with
$\nu_{{p}}= 1/(d+2)$ ($d=3$) which describes the scaling of the chain in the limit $N/ \tilde\ell_{p} \gg 1$ \cite{Nakanishi} 
and is correctly recovered by Eq.~\!(\ref{end_to_end_distance}). In the opposite stiff or rod-like chain limit of 
{$N / \tilde{\ell}_{{p}} \ll 1$},
Eq.~\!(\ref{end_to_end_distance}) recovers the trivial result that $\tilde{R}_N= N$.
The quantity
$\nu=0.588$ is the Flory exponent, and $A=0.8$, 
$a_1=0.1$ and $b_1=0.9$ are constants.
In the intermediate regime $4 \lesssim N/\tilde{\ell}_{{p}} \lesssim 400$ which here corresponds to 
$10^2 \lesssim N  \lesssim 10^4$ for $\tilde{\ell}_{{p}}=25$, a crossover occurs from a rod-like chain
to a Gaussian (ideal) polymer, followed by an eventual crossover to a self-avoiding chain for 
$N/\tilde{\ell}_{{p}} \gg 10^6$ \cite{Hsu} as 
can be seen in the inset of 
Fig.~\!\ref{fig:end_to_end_distance}.
Remarkably, we find that Eq.~\!(\ref{end_to_end_distance}) is universally valid with the same values of $A, a_1$ and $b_1$
for a wide range of values of $\tilde{\ell}_p$, as shown in SM. 
It should be noted that the amplitude $A$ is fixed
by the equilibrium scaling of the chain, and thus only $a_1$ and $b_1$ are fitting parameters.

{\it (c) Time evolution of the tension front} --
Using $\tilde{R} (\tilde{t})$ in Eq.~\!(\ref{end_to_end_distance}) together with the mass conservation 
$N = \tilde{l} + \tilde{s}$, where $\tilde{l} = \tilde{R}$, the equation of motion for the tension front 
in the TP stage for the SS regime (see Fig.~\!\ref{fig:schimatic_1}(a)) can be derived as
\begin{equation}
\dot{\tilde{R}} (\tilde{t}) \!=\!
\frac{ \tilde{\phi} (\tilde{t}) ~ (\cal{G}+\cal{H}) }{  2 \tilde{R} (\tilde{t}) - (\cal{G}+\cal{H}) },
\label{evolution_of_R_propagation_SS}
\end{equation}
where 
\begin{eqnarray}
{\cal{G}} &=& \!\!  \frac{ \tilde{R}^2_{F} }{ N } 
\bigg[ 2\nu -(2-2\nu) ~ \exp \bigg( -\frac{2 a_1 N^{2}}{ \tilde{R}^2_{F} } \bigg) \bigg] \nonumber\\
&+& \frac{(4\nu-2) \tilde{R}^4_{F} }{2 a_1 N^3 } 
\bigg[ -1 + \exp \bigg( -\frac{2 a_1 N^{2}}{ \tilde{R}^2_{F} } \bigg) \bigg], \nonumber\\
{\cal{H}} &=& 2 ~ \tilde{\ell}_{{p}} \bigg[ 1 - \exp \bigg( -\frac{b_1N}{ \tilde{\ell}_{{p}} } \bigg) \bigg].
\label{G_evolution_of_R_propagation_SS_2}
\end{eqnarray}

\begin{figure}[b]
\begin{center}
    \includegraphics[width=0.38\textwidth]{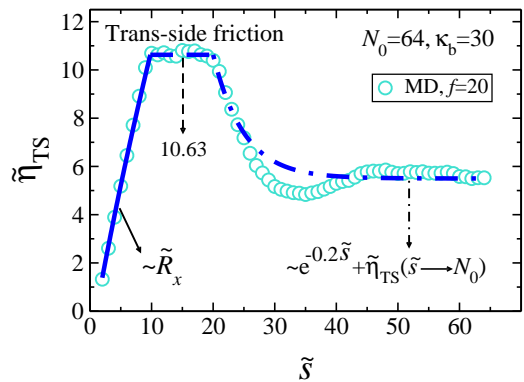}
\caption{
The {\it trans} side friction $\tilde{\eta}_{_\textrm{TS}}(\tilde{s})$ as a function of $\tilde{s}$ for fixed values 
of the chain length $N_0 = 64$, bending rigidity $\kappa_{\textrm{b}}=30$, and external 
driving force $f=20$. The turquoise circles are MD data. The blue solid, dashed and dashed-dotted lines
represent the three different regimes (see the text and SM for details).
}
\label{fig:trans-side-friction}
\end{center}
\end{figure}

In the PP stage (see Fig.~\ref{fig:schimatic_1}(b)) 
the correct closure relation is $\tilde{l}+\tilde{s}=N_0$. Then one 
can derive the equation of motion for the tension front in PP stage as
\begin{equation}
\dot{\tilde{R}} (\tilde{t}) = - \tilde{\phi} (\tilde{t}).
\label{evolution_of_R_post_propagation_SS}
\end{equation}

To find the solution, in the TP stage, Eqs.~\!(\ref{BD_equation_1}), (\ref{phi_equation}), (\ref{Gamma_equation}) and 
(\ref{evolution_of_R_propagation_SS}) must be solved self-consistently 
while in the PP stage, Eqs.~\!(\ref{BD_equation_1}), (\ref{phi_equation}), 
(\ref{Gamma_equation}), (\ref{evolution_of_R_post_propagation_SS}) must be solved.

{\it Results: (a) Trans side friction} -- 
We present the waiting time distribution $w(\tilde{s})$, which is the time that each bead spends at the pore,
in SM. 
The data clearly show that in order to have a quantitative theory, we must include 
$\tilde{\eta}_{\rm TS}(t)$ in Eq.~\!(\ref{Gamma_equation}).

It is expected that the {\it trans} side friction is a complicated function of the driving force, chain length and
the bending rigidity, and the present IFTP theory does not allow us to derive it analytically. To this end, we have
extracted it numerically from the MD simulations as shown in
Fig.~\!\ref{fig:trans-side-friction}. Details and additional data for smaller driving forces and for different persistence lengths are 
in SM. 
We can identify three distinct
regimes in $\tilde{\eta}_{\rm TS}(\tilde{s})$. For small $\tilde{s}/N_0$, we find that the friction grows proportional to the $x$ component
of the end-to-end distance $\tilde{R}_x$. After this initial stage it saturates to a constant value (here $\approx 10.63$), which from the MD simulations
indicates buckling of the {\it trans} part of the chain. This buckling of the chain reduces the friction and we find an approximately 
exponential decay of the friction towards an asymptotic constant $\tilde{\eta}_{\rm TS}(N_0) \approx 5.5$.

{\it (b) Translocation time exponent} --
The scaling of the average translocation time as a function of the chain length $\tau \propto N_0^{\alpha}$ is
a fundamental characteristic of translocation dynamics. For flexible chains it scales
as $\tau = a_p N_0 + a_c N_0^{\nu +1}$, where $a_p$ and $a_c$ are constants. The first term is due
to the pore friction which causes a significant finite-size correction to the asymptotic scaling 
where $\alpha = {\nu +1}$ \cite{ikonen2012a,ikonen2012b,ikonen2013,jalal2014,jalal2015}.
The asymptotics is, of course, recovered for the semi-flexible chains in the large $N_0$ limit 
when $\tilde{\ell}_p/N_0 \ll 1$.
On the other hand, in the limit of a rod-like polymer $\tau \propto N_0^2$. 
Following Ref.~\!\cite{jalal2014}, we can derive an analytic expression for $\tau$ 
by assuming that only the external driving force 
$\tilde{f}$ contributes to the total force in the BD equation~\eqref{BD_equation_1}. This leads to reduction of 
Eq.~\!\eqref{phi_equation} to $\tilde{\phi}(\tilde{t})= \tilde{f} / \big[ \tilde{R}(\tilde{t}) + \tilde{\eta}_{\textrm{p}} + 
\tilde{\eta}_{_\textrm{TS}} \big]$, and the total translocation time can be written as
\begin{equation}
\tilde{\tau} = \frac{1}{\tilde{f}} \bigg[ \int_0^{N_0} \tilde{R}_N dN 
+ \tilde{\eta}_{\textrm{p}} N_0 \bigg] + \tilde{\tau}_{\textrm{TS}},
\label{scaling_trans_time_SS_2}
\end{equation}
\begin{figure}[t]
\begin{center}
    \includegraphics[width=0.38\textwidth]{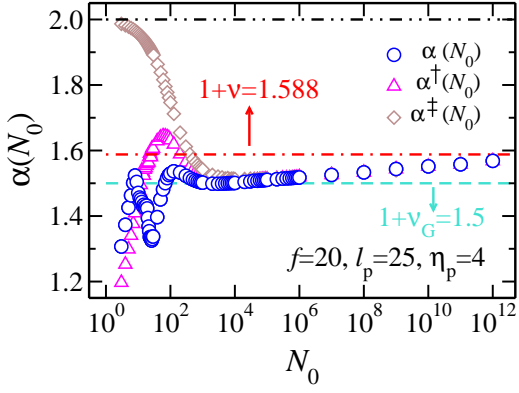}
\caption{The effective translocation time exponents as a function of the 
chain length $N_0$ for $\ell_{{p}}=25$ and pore friction $\eta_{\textrm{p}}=4$, and external driving force $f=20$. 
The blue circles show the translocation exponent $\alpha$ as a function of $N_0$,
while pink triangles and brown diamonds show the rescaled translocation exponents $\alpha^{\dag}$ and 
$\alpha^{\ddag}$, respectively. The horizontal black dashed-dotted-dotted, 
red dashed-dotted and turquoise dashed lines show the asymptotic rod-like, excluded volume chain 
and Gaussian scaling limits, respectively. See text for details.}
\label{fig:trans-exponent_1}
\end{center}
\end{figure}
where $\tilde{\tau}_{\textrm{TS}} = \big[ \int_0^{N_0}\tilde{\eta}_{_\textrm{TS}} dN + 
\int_0^{\tilde{R}_{N_0}} (\tilde{\eta}_{_\textrm{TS,pp}} - \tilde{\eta}_{_\textrm{TS,tp}}) d\tilde{R} \big]/\tilde{f}$ 
is the {\it trans} side contribution to the translocation time.
The second term in $\tilde{\tau}_{TS}$ is due to non-monotonic behavior of $\tilde{\eta}_{\textrm{TS}}$ 
in the TP and PP stages.
In the rod limit we obtain the simple analytical result that
\begin{equation}
\tilde{\tau} = \frac{1}{\tilde{f}} \bigg[ \tilde{\eta}_{{p}} N_0 + N_0^2 \bigg] ,
\label{scaling_trans_time_Rod_like_2}
\end{equation}
which gives the asymptotic exponent $\alpha = 2$.
The corresponding effective exponents will be between unity and two.

\begin{figure}[b]
\begin{center}
    \includegraphics[width=0.38\textwidth]{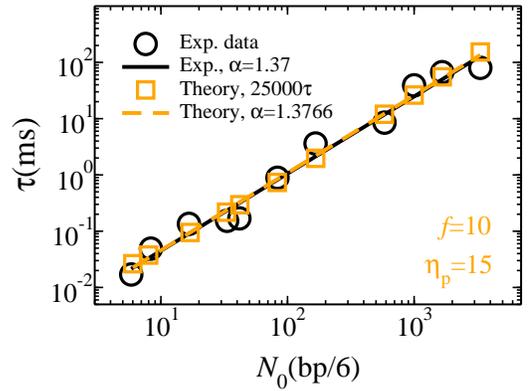}
\caption{Translocation time $\tau$ as a function of the chain length $N_0$. 
Black circles are experimental data in Fig.~\!6(c) of Ref.~\!\cite{Spencer2014} 
while orange squares are data from the IFTP theory, 
where we have used the coarse grained model. Each bead contains 6 bps. The value of the 
external driving force in the IFTP theory is $f=10$ and the pore friction is $\eta_{\textrm{p}}=15$. 
The translocation time for the IFTP theory has been multiplied by a factor of 25000 to agree with
the experimental time scale. 
The black solid and orange dashed lines are linear fitting curves to experimental and IFTP theory, respectively.
Similar results can be obtained for the values of the external driving forces $f=5$ and $20$.
Details on mapping the experimental data to theory are explained
in Sec.~\!I of SM. 
} 
\label{fig:trans-exponent}
\end{center}
\end{figure}

To quantify the influence of the {\it trans} side and pore friction on the effective translocation exponent
we define two rescaled translocation exponents $\alpha^{\dag}$ and $\alpha^{\ddag}$ 
as
$\tau^{\dag}= \tau - \tau_{\textrm{TS}} \sim N_0^{\alpha^{\dag}} $ and 
$\tau^{\ddag}= \tau - \tau_{\textrm{TS}} - a_{\textrm{p}} N_0 \sim N_0^{\alpha^{\ddag}} $, 
respectively.
In the short ($N_0/\tilde{\ell}_{{p}} \lesssim 4$) and intermediate ($4 \lesssim N_0/ \tilde{\ell}_{{p}} \lesssim 400$) chain limits, 
contributions from both the {\it trans} side and pore friction are important as can be seen in 
Eq.~\!(\ref{scaling_trans_time_SS_2}). 

In Fig.~\!\ref{fig:trans-exponent_1} we show the detailed dependence of the effective 
translocation time exponents as a function of the chain length 
$N_0$ for constant values of the persistence length $\ell_{{p}}=25$, pore friction $\eta_{\textrm{p}}=4$ and 
driving force $f=20$. The blue circles show the effective value of the total 
$\alpha$ as a function of $N_0$. The non-monotonic behavior of the {\it trans} side friction leads into a non-monotonic
dependence of $\alpha$ on $N_0$. Interestingly enough, there is an extended intermediate range of chain lengths
where the exponent is very close to the Gaussian value 
$\alpha= 3/2$ and slowly approaches its
asymptotic value of $1+\nu=1.588$ from below. We note that in order to see this crossover it is necessary to have
a full scaling form for the end-to-end distance of the form of Eq.~\!(\ref{end_to_end_distance}).

To quantify how the {\it trans} side friction affects the effective translocation exponent, in Fig.~\!\ref{fig:trans-exponent_1} 
we plot $\alpha^{\dag}$ (pink triangles). It approaches $\alpha$ for $N_0 > 10^4$, where the {\it trans} side friction becomes negligible. 
Finally, the rescaled translocation exponent $\alpha^{\ddag}$ (brown diamonds), which is the effective translocation time exponent 
in the absence of both {\it trans} side and pore friction, is also plotted as a function of $N_0$. This exponent recovers the rod-like 
limit for very short chains. It merges with the other two effective exponents to the almost Gaussian value at intermediate lengths
and eventually approaches $\nu +1$, as expected.

Finally, we compare the results of IFTP theory with relevant experiments. In Fig.~\!\ref{fig:trans-exponent}, we present the
translocation time obtained from experiments (black circles) and from the augmented IFTP theory (orange squares) as a function of 
the chain length $N_0 ({\textrm{bp}}/6)$, for fixed values of external driving force $f=10$ and pore friction $\eta_{\textrm{p}}=15$.
The value of external driving force $f=10$ corresponds to potential difference $V=200$ mV across the pore in the experiments \cite{Spencer2014}
(for more information see SM.) 
To match the length scales, we coarse grain such that one bead in our model contains 6 bps. With this choice the translocation exponent 
from the IFTP theory (orange dashed line) is in good agreement with the exponent from the experimental data (black solid line).

{\it Summary and Conclusions} --
We have shown here that in addition to the case of fully flexible polymers, 
the IFTP theory provides the proper theoretical framework for driven translocation
of semi-flexible polymers. The two key quantities required are an explicit determination of the {\it trans}
side friction and a proper analytical formula for the end-to-end distance of semi-flexible polymers. The augmented
IFTP theory can quantitatively describe all the relevant scaling regimes for the scaling exponent of the average translocation
time, and crossover between them. It also reproduces the exactly known limits and is in good agreement with available
experimental data. 
 
{\it Acknowledgments} -- J.S. thanks V. Thakore, H.-P. Hsu and R.R. Netz for enlightening discussions. This work was supported by the Academy 
of Finland through its Centers of Excellence Program under Project Nos. 251748 and 284621. The numerical 
calculations were performed using computer resources from the Aalto University School of Science ``Science-IT'' project, and
from CSC - Center for Scientific Computing Ltd. This research was funded in part by the National Institutes of Health (R01-HG009186, M.W.).


\end{document}


\title{Supplementary Material for\\ ``Driven translocation of a semi-flexible polymer through a nanopore''}

\author{Jalal Sarabadani}
\email{jalal.sarabadani@aalto.fi}
\affiliation{Department of Applied Physics and COMP Center of Excellence, Aalto University School of Science, 
P.O. Box 11000, FI-00076 Aalto, Espoo, Finland}

\author{Timo Ikonen}
\affiliation{~VTT Technical Research Centre of Finland Ltd., P.O. Box 1000, FI-02044 VTT, Finland}

\author{Harri M\"okk\"onen}
\affiliation{Department of Applied Physics and COMP Center of Excellence, Aalto University School of Science, 
P.O. Box 11000, FI-00076 Aalto, Espoo, Finland}

\author{Tapio Ala-Nissila}
\affiliation{Department of Applied Physics and COMP Center of Excellence, Aalto University School of Science, 
P.O. Box 11000, FI-00076 Aalto, Espoo, Finland}
\affiliation{Department of Physics, Box 1843, Brown University, Providence, Rhode Island 02912-1843.}

\author{Spencer Carson}
\affiliation{Department of Physics, Northeastern University, Boston MA 02115}

\author{Meni Wanunu}
\affiliation{Department of Physics, Northeastern University, Boston MA 02115}
\maketitle

\date{December 22, 2016}

\section{Molecular dynamics model} \label{MD}

In our Molecular Dynamics (MD) simulations the polymer is modeled by a bead-spring chain \cite{Kremer}. The excluded volume interaction 
between the beads is given by the repulsive Lennard-Jones (LJ) potential 
$U_{\textrm{LJ}} = 4\epsilon [(\frac{\sigma}{r})^{12} - (\frac{\sigma}{r})^{6} ] +\epsilon$ for $r \leq 2^{1/6} \sigma$,
and zero for $r > 2^{1/6} \sigma$, where $\epsilon$ is the depth of the potential well, $\sigma$ is the diameter 
of each bead, and $r$ is the distance between the beads. We use the finitely extensible nonlinear elastic (FENE) spring 
interaction to interconnect neighboring beads, given by $U_{\textrm{FENE}} = -\frac{1}{2} k R_0^2 \ln \big( 1- r^2 /R_0^2 \big)$, 
where $k$ is the spring constant and $R_0$ is the maximum allowed distance between consecutive beads.
We introduce the stiffness of the chain by adding an angle dependent cosine potential
$U_{\textrm{bend}} (\theta_i) =  \kappa_{{b}} (1 - \cos \theta_i) $ between successive bonds, 
which connect $(i-1)^{\rm th}$ and $i^{\rm th}$, and the $i^{\rm th}$ and $(i+1)^{\rm th}$ beads, where the bending rigidity 
$\kappa_{{b}}$ is the interaction strength.

The physical wall is constructed by using the repulsive LJ interaction 
$U_{\textrm{LJ}} = 4\epsilon [(\frac{\sigma}{x})^{9} - (\frac{\sigma}{x})^{3} ]$, where $x$ is the coordinate in
the direction perpendicular to the wall. The region of space with $x<0$ is called the {\it cis} side and with $x>0$ is the 
{\it trans} side. To construct the pore, 16 beads with diameter of $\sigma$ are placed on
a circle with diameter of $d=3\sigma$. The center of the pore is at $x=0$ and the pore is parallel to the wall. 
The thickness of the pore is $\sigma$ and the interaction between monomers 
and the pore particles is repulsive LJ with the same parameters as of the excluded volume interactions between the polymer beads.
The external driving force, $f$, which is in the positive $x$ direction, only acts to the beads that are inside the 
pore. 

Using Langevin dynamics the equation of motion 
for the $i$th bead is written as 
$m \ddot{r}_i = - \nabla ( U_{\textrm{LJ}}  + U_{\textrm{FENE}} + U_{\textrm{bend}} + U_{\textrm{ext}} ) - \eta v_i + \xi_i$.
Here, $m$ in the mass of each monomer, $\eta$ is the friction coefficient of the solvent, $v_i$ is the monomer velocity,
and $\xi_i$ is an uncorrelated random force with $\langle \xi_i (t) \xi_j (t') \rangle = 2 \eta k_B T \delta_{i,j} \delta (t-t')$. 
By using LJ units, the mass of each bead is chosen as $m=1$, the length is expressed in the unit of $\sigma$, and
the unit of time is $\sigma \sqrt{m/\epsilon}$. Temperature $T$ is measured in units of $\epsilon / k_B$,
and the unit of energy is $\epsilon= k_B T$. In LJ units the parameters of the interactions potential, length,
mass, spring constant, maximum allowed distance between consecutive beads, bending rigidity, and friction coefficient have been 
chosen as $\epsilon=1$, $\sigma=1$, $m=1$, $k=30$, $R_0 =1.5 \sigma$ $\kappa_b = 30$, and $\eta =0.7$, respectively, and the external driving 
force as $f=5, 10$ and $20$. Here, $k_B T = 1.2$.

In our simulations, we have used the coarse grained bead-spring model. 
According to the relation $\ell_{{p}}= \kappa_{{b}}/(k_B T)$ in 3D, with the value of $\kappa_{{b}}=30$, 
the persistence length is $\ell_{{p}} = 25$. As the persistence length of DNA is 150 bps, 
in our model each bead corresponds approximately to 6 bps. The mass of a bead is about 3744
amu while its size is chosen as $\sigma = 2$ nm,
and the interaction strength is $3.39 \times 10^{-21}$J at room temperature ($T = 295$ K). 
Therefore, the time scale in LJ unit is 85.6 ps. By assuming the effective charge of 0.094 e for each
unit charge \cite{Branton_PRL_2003,Meller_BioPhysJ_2004}, twelve unit charges per bead and with a force
scale of 2.0 pN, an external driving force of $f=10$ corresponds to a voltage of 200 mV across the pore.

In the beginning of the translocation process, first we fix the first bead (head of polymer chain) 
at the pore and equilibrate the system in the {\it cis} side, after which we start the actual translocation by turning on the external 
driving force and releasing 
the first bead at $t=0$. The translocation time $\tau$ is defined as the time when the last bead of the chain enters 
to the {\it trans} side. It is important to note that reflective boundary conditions must not be used for the chain, but in the
case the chain escapes from the pore to the {\it cis} side, the translocation must be re-started from a new equilibrium configuration at $t=0$.

\section{End-to-end distance formula} \label{ETED}

We propose the following semi-analytic expression for the end-to-end distance of a semi-flexible polymer chain with contour length $N$ and persistence length 
$\tilde{\ell}_p$:
\begin{equation}
\tilde{R}_N = \bigg\{ \tilde{R}^2_{F}
- \frac{\tilde{R}^4_{F} }{2 a_1 N^2 } \bigg[ 1- \exp \bigg( - \frac{2 a_1 N^2 }{ \tilde{R}^2_{F} }
\bigg) \bigg] + 2 \tilde{\ell}_{{p}} N \!
- \frac{2 \tilde{\ell}_{{p}}^{2}  }{b_1}  
\bigg[ 1 \!-\! \exp \bigg( \!\!- \frac{b_1 N }{ \tilde{\ell}_{{p}}} \bigg) \! \bigg] \! \bigg\}^{\frac{1}{2}},
\label{end_to_end_distance}
\end{equation}
where $A=0.8$ and $\tilde{R}_{F}= A \tilde{\ell}_{{p}}^{\nu_{{p}}} N^{\nu}$,  with
$\nu_{{p}}= 1/(d+2)$ (here $d=3$). Equation \!(\ref{end_to_end_distance}) correctly recovers the scaling of the fully flexible 
self-avoiding chain in the limit $N/ \tilde\ell_{p} \gg 1$ \cite{Nakanishi} 
as $\tilde{R} (N/ \tilde\ell_{p} \gg 1) = \tilde{R}_{F}= A \tilde{\ell}_{{p}}^{\nu_{{p}}} N^{\nu}$. 
In the opposite stiff or rod-like chain limit of $N / \tilde{\ell}_{{p}} \ll 1$, Eq.~\!(\ref{end_to_end_distance}) 
gives the end-to-end distance as $\tilde{R}_N= \sqrt{a_1 +b_1} N$, where by setting $a_1 +b_1 = 1$ 
(e.g. $a_1= 0.1$ and $b_1=0.9$) we recover the trivial result that $\tilde{R}_N=  N$. 
In the intermediate regime $N/ \tilde\ell_{p} \sim 10^2$, the end-to-end distance is obtained from Eq.~\!(\ref{end_to_end_distance}) 
as $\tilde{R}_N= 2 \tilde{\ell}_{{p}} N$ which is a characteristics of the Gaussian chain.

\begin{figure}[t]
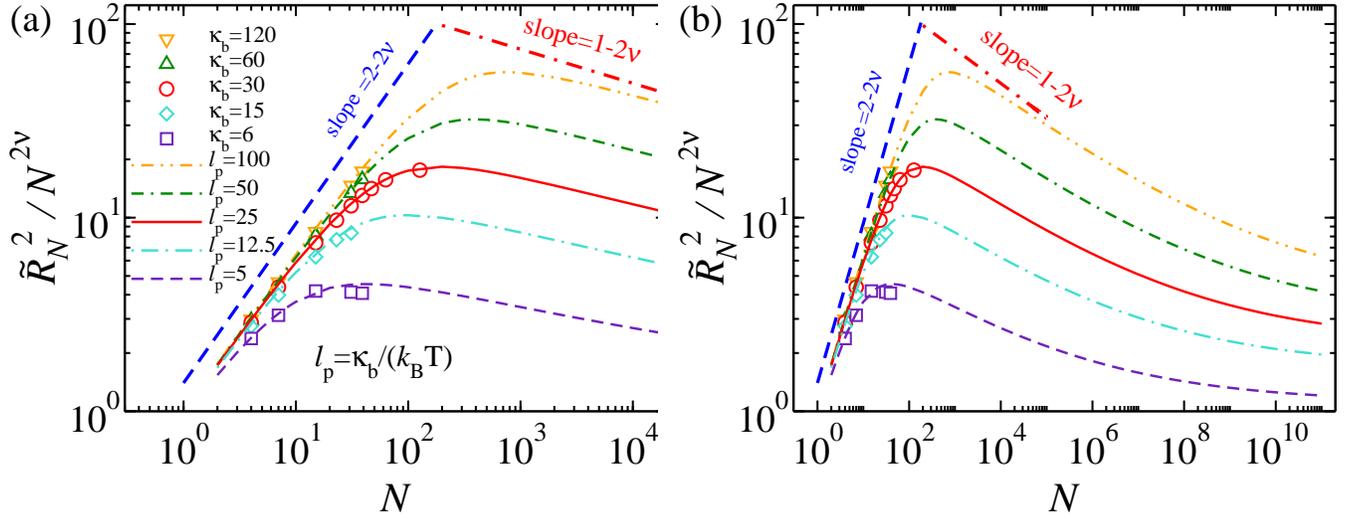

\begin{center}
    \includegraphics[width=0.49\textwidth]{figure1a_supp.eps}
    \includegraphics[width=0.49\textwidth]{figure1b_supp.eps}
\caption{
(a) Normalized end-to-end distance $\tilde{R}_N^2/N^{2\nu}$ as a function of the contour length of the polymer $N$
for $k_B T=1.2$ and various values of the bending rigidity (in the MD simulations): $\kappa_{{b}}= 6$ (purple squares), 
15 (turquoise diamonds), 30 (red circles), 60 (green upward triangles) and 120 (orange downward triangles), which correspond 
to $\ell_p= 5$ (purple dashed line), 12.5 (turquoise dashed-dotted line), 25 (red solid line), 50 
(green dashed-dashed-dotted line) and 100 (orange dashed-dotted-dotted line), respectively, according to 
$\ell_{{p}}= \kappa_{{b}}/(k_B T)$ in 3D.
The lines are from the analytical formula of Eq.~\!(\ref{end_to_end_distance}). (b) An extended range of $N$ shows
how the scaling of $\tilde{R}_N$ eventually crosses over to that of a self-avoiding chain at very large $N/\tilde{\ell}_p$ from 
the intermediate range Gaussian behavior. In the MD simulations the chain lengths are $N=5, 8, 16, 24, 32, 40, 64$ and 128 for 
$\kappa_b = 30$, and $N= 5, 8, 16, 24$ and 32 for $\kappa_b = 6, 15, 60$ and 120.
} 
\label{fig:end_to_end_distance}
\end{center}
\end{figure}

To show the validity of the expression for the end-to-end distance, 
in Fig.~\!\ref{fig:end_to_end_distance} we compare results from Eq.~\!(\ref{end_to_end_distance})
with MD simulations by 
presenting the normalized end-to-end distance $\tilde{R}_N^2/N^{2\nu}$ as a function of the chain length $N$ for
various values of the bending rigidity $\kappa_{{b}}= 6$ (purple squares), 
15 (turquoise diamonds), 30 (red circles), 60 (green upward triangles) and 120 (orange downward triangles) which correspond 
to $\ell_p= 5$ (purple dashed line), 12.5 (turquoise dashed-dotted line), 25 (red solid line), 50 
(green dashed-dashed-dotted line) and 100 (orange dashed-dotted-dotted line), respectively. As can be seen,
Eq.~\!(\ref{end_to_end_distance}) is able to reproduce the end-to-end 
distance of semi-flexible polymers with different persistence lengths for a wide range of chain parameters. In particular,
in the intermediate regime between the stiff rod and fully flexible self-avoiding chains Eq.~\!(\ref{end_to_end_distance}) 
correctly describes the Gaussian behavior.

\section{{Trans} side friction}\label{Trans_side_friction}

\begin{figure*}[t]
\begin{center}
    \begin{minipage}[b]{0.49\textwidth}
        \includegraphics[width=0.7\textwidth]{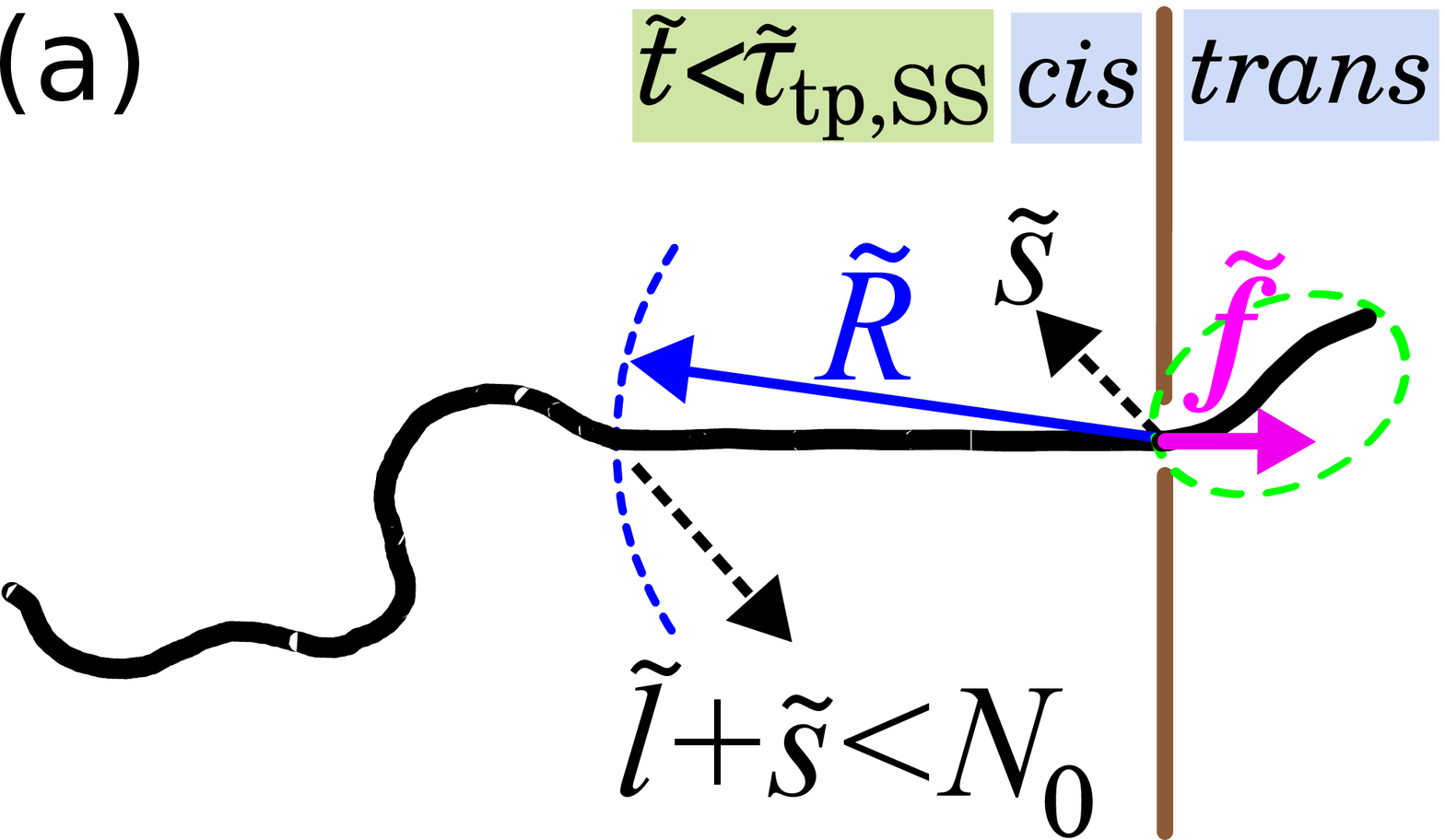}
    \end{minipage} \hskip-1.0cm
    \begin{minipage}[b]{0.4\textwidth}
        \includegraphics[width=0.7\textwidth]{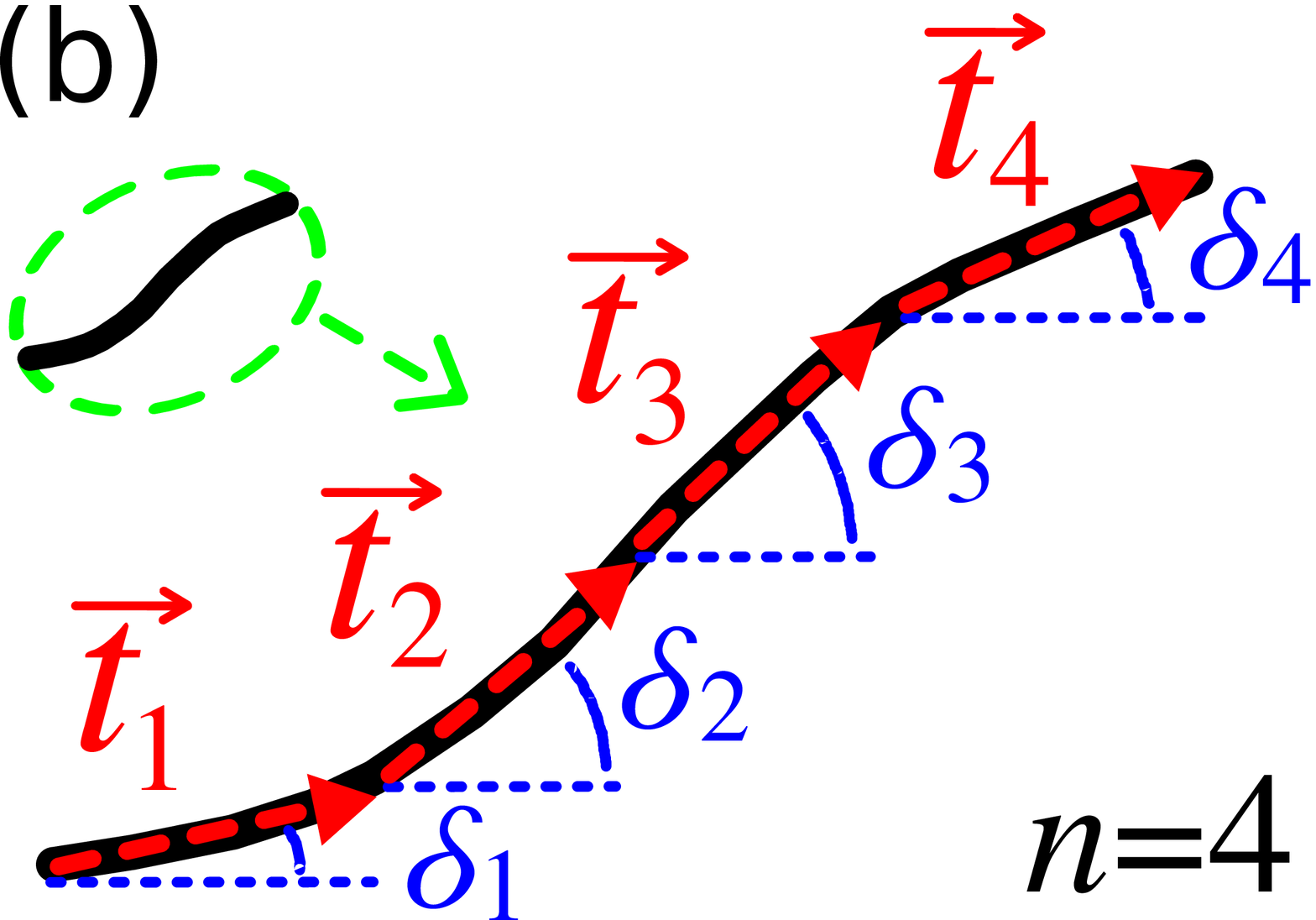}
    \end{minipage} \hskip+0cm
\caption{
(a) A schematic of the translocation process in the tension propagation stage, i.e. $\tilde{t} < \tilde{t}_{\textrm{tp,SS}}$, 
for the strong stretching regime. 
(b) A schematic representation of the chain in the {\it trans} side when
$\tilde{s} = 5$, in the green dashed ellipsoid in the {\it trans} side of panel (a). The tangential vector $\vec{t}_i$ 
connects beads with translocation coordinates $\tilde{s}_i$ and $\tilde{s}_{i+1}$ and therefore here the number of tangential 
vectors on the {\it trans} side is $n=4$. The angle between $\vec{t}_i$ and the direction of the external driving force $\tilde{f}$, 
which is $\hat{x}$, is denoted by $\delta_i$.
} 
\label{fig:schimatic_1}
\end{center}
\end{figure*}

\begin{figure*}[t]
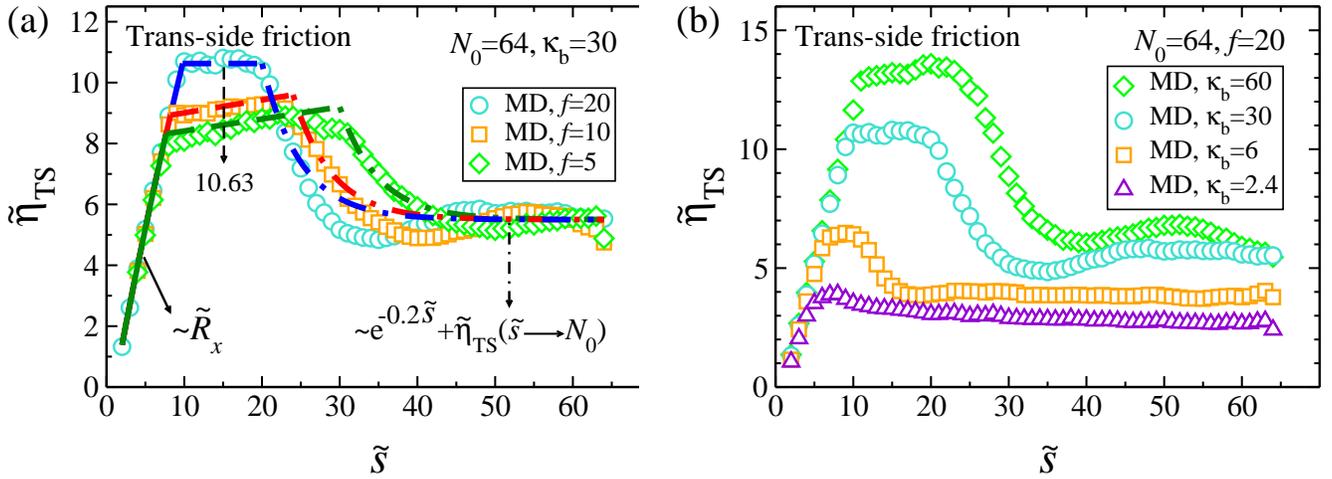

\begin{center}
    \begin{minipage}[b]{0.49\textwidth}
        \includegraphics[width=0.98\textwidth]{figure3a_supp.eps}
    \end{minipage} \hskip0.0cm
    \begin{minipage}[b]{0.49\textwidth}
        \includegraphics[width=0.98\textwidth]{figure3b_supp.eps}
    \end{minipage} \hskip+0cm
\caption{
(a) The {\it trans} side friction $\tilde{\eta}_{_\textrm{TS}}$ as a function of the translocation coordinate $\tilde{s}$ for 
chain length $N_0 = 64$, bending rigidity coefficient $\kappa_{{b}}=30$ and various values of the external 
driving force $f=5, 10$ and $20$. The green diamonds ($f=5$), orange squares ($f=10$) and turquoise circles ($f=20$)
are MD data.
For $f=20$, the blue solid line represents the {\it trans} side friction at the beginning of 
the translocation process, which is proportional to the $x$ component of the end-to-end distance. 
The horizontal blue dashed line shows that the {\it trans} side friction has a constant value of 
$\approx 10.63$ during the first buckling stage. Finally, the blue dashed-dotted line exhibits the {\it trans} 
side friction after the buckling has already occurred, demonstrating an exponential decay
to the asymptotic value of the {\it trans} side friction, $\tilde{\eta}_{_\textrm{TS}} (\tilde{s} \rightarrow N_0)$.
The green and red lines represent these approximate analytical fits for the {\it trans} side friction for $f=5$ and $f=10$, 
respectively. (b) $\tilde{\eta}_{_\textrm{TS}}$ as a function of $\tilde{s}$ for 
chain length $N_0 = 64$,  external driving force $f=20$ and various values of the bending rigidity coefficient 
$\kappa_{{b}}=2.4, 6, 30$ and 60. 
The green diamonds ($\kappa_b=60$), orange squares ($\kappa_b=30$), turquoise circles ($\kappa_b=6$) and violet triangles 
($\kappa_b=2.4$) are MD data.
} 
\label{fig:trans-side-friction-S1}
\end{center}
\end{figure*}

The {\it trans} side friction discussed in the main article has complicated dependence on the physical parameters of the
translocation process. We have
extracted it numerically from the MD simulations by calculating the normalized angular cosine-correlation function 
$C(n)= \cos \delta_{1} \cos \delta_{2} \ldots \cos \delta_{n}/ \cos \delta_{1}$ (cf. Fig.~\!\ref{fig:schimatic_1}(b)) 
for each integer $\tilde{s}=n+1$. To estimate the friction on the {\it trans} side, we define a cut-off value $n^*$ 
for the correlation function such that $C(n^*)=1/e$. The actual contribution to the friction in given by the values 
of $C(i)<C(n^*)$. Then, the {\it trans} side friction for the given $\tilde{s}$ is written as 
$\eta_{\textrm{TS}} (\tilde{s}) = \sum_{i=1}^{n^*} \cos \delta_i$.
In Fig.~\!\ref{fig:trans-side-friction-S1}(a) we show the numerically extracted {\it trans} side friction as a function 
of the translocation coordinate $\tilde{s}$ for fixed chain length of $N_0 = 64$, bending rigidity $\kappa_b = 30$ and 
for different values of the external driving force $f=5, 10$ and $20$. 
In Fig.~\!\ref{fig:trans-side-friction-S1}(b) the same quantity as in 
Fig.~\!\ref{fig:trans-side-friction-S1}(a) is presented as a function of $\tilde{s}$ but for a fixed value of $f=20$ and different 
values of the bending rigidity $\kappa_b = 2.4, 6, 30$ and 60. We can identify three distinct regimes in $\tilde{\eta}_{\rm TS}(\tilde{s})$. 
For small $\tilde{s}/N_0$, we find that the friction grows proportional to the $x$ component
of the end-to-end distance $\tilde{R}_x$. After this initial stage it saturates to a constant value (for example 10.63 for $f=10$), 
which from the MD simulations
indicates buckling of the {\it trans} part of the chain. This buckling of the chain reduces the friction and we find an approximately 
exponential decay of the friction towards another constant value $\tilde{\eta}_{\rm TS}(N_0) \approx 5.5$ 
(see Fig.~\!\ref{fig:trans-side-friction-S1}).
There is currently no analytic formula available for $\tilde{\eta}_{\rm TS}$.

As explained above and also in the main text of the article, there are three regimes 
for the trans side friction. Here, we elaborate on the physical mechanisms behind these regimes.
According to our MD simulations
at the early stages of the translocation process  $\tilde{s} / \tilde{\ell}_p \ll 1$ the trans side chain is rod-like.
Therefore, the trans side friction 
increases roughly linearly. At intermediate times where $\tilde{s} / \tilde{\ell}_p = {\cal O}(1)$, the chain has advanced
far enough such that the trans side starts to bend due to fluctuations and increased friction of the solvent. In this regime 
the friction saturates to an intermediate value, which becomes larger for either increasing driving force (cf. Fig.~\!3(a)) or stiffness (cf. Fig.~\!3(b), see also
Ref. \cite{DaiPolymers2016} where similar behavior has been observed). In the late stages of translocation where
$\tilde{s} / \tilde{\ell}_p \gg 1$, the trans side friction approaches its asymptotic constant value 
$\tilde{\eta}_{TS} (\tilde{s} \rightarrow N_0)$. As can be seen in Fig. 3(b), in the limit
of fully flexible chains the asymptotic
constant value is rapidly attained and can thus be incorporated in a constant, effective pore friction as we
have already previously shown \cite{jalal2014}.

\section{Waiting time distribution}\label{Waiting_time}
\begin{figure}[t]
\begin{center}
    \includegraphics[width=0.49\textwidth]{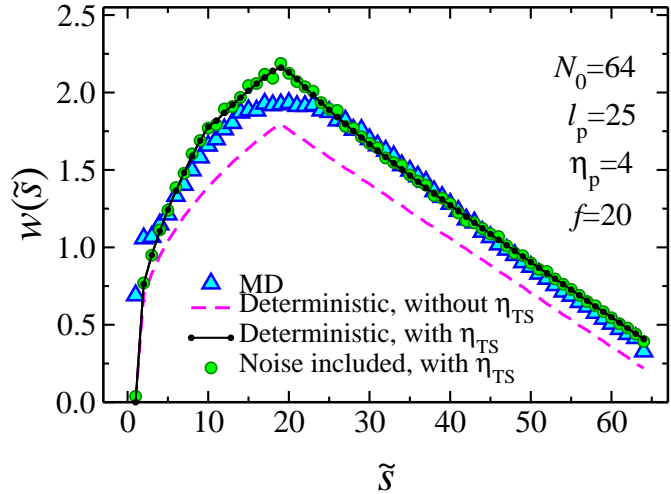}
\caption{
The waiting time distribution $w (\tilde{s})$ as a function of the translocation coordinate $\tilde{s}$. 
The parameters here are chain length $N_0=64$, persistence length 
$\ell_{{p}}=25$, pore friction $\eta_{\textrm{p}}= 4$, and the external driving force $f=20$.
The MD simulation data are presented by blue triangles.
The pink dashed curve is the waiting time when the {\it trans} side friction is not explicitly taken into account \cite{jalal2014}.
The solid black line is the result from the IFTP theory with $\tilde{\eta}_{\rm TS}$, and the green circles represent the 
waiting time when noise is added to the equation of motion \cite{jalal2014}.
} 
\label{fig:waiting_time}
\end{center}
\end{figure}

In Fig.~\!\ref{fig:waiting_time} we show the waiting time distribution $w(\tilde{s})$, 
which is the time that each bead spends at the pore, as obtained from the MD simulations (blue triangles). 
The pink dashed line is the result obtained from the previous IFTP theory of Ref.~\!\cite{jalal2014}
by assuming that the {\it trans} side friction is implicitly included in $\eta_{\rm p} = const.$, 
which is an excellent approximation for the fully flexible chains. 
The data clearly show that in order to have a quantitative theory, we must include 
$\tilde{\eta}_{\rm TS}(t)$ in Eq.~\!(3) of the main article.

\section{Translocation time exponent}\label{Translocation_exponent}

\begin{figure}[t]
\begin{center}
    \includegraphics[width=0.49\textwidth]{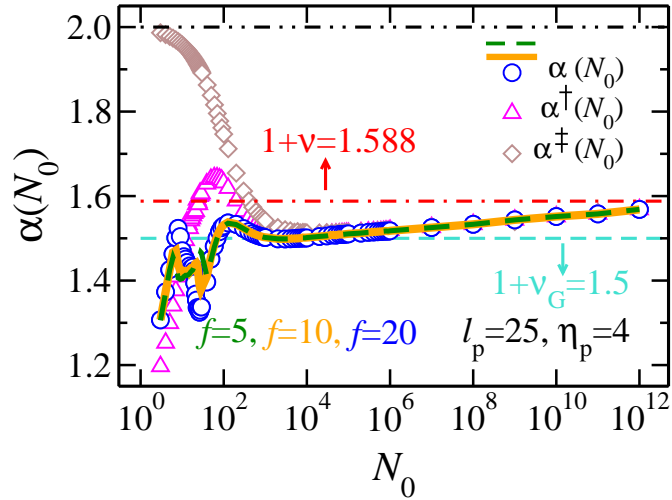}
\caption{The effective translocation time exponent $\alpha$ as a function of the chain length $N_0$ for
persistence length $\ell_{{p}}=25$ and pore friction $\eta_{\textrm{p}}=4$, for various values of the external 
driving force $f=5$ (green dashed line), $10$ (orange solid line) and $20$ (blue circles).
The pink triangles and brown diamonds show the rescaled translocation exponents $\alpha^{\dag}$ and 
$\alpha^{\ddag}$, respectively, as a function of $N_0$. The horizontal black dashed-dotted-dotted, 
red dashed-dotted and turquoise dashed lines show the asymptotic rod-like, exculed volume and the Gaussian chain
limits, respectively.
}
\label{fig:trans-exponent_1_S1}
\end{center}
\end{figure}

In Fig. \ref{fig:trans-exponent_1_S1} the effective translocation time exponent $\alpha$ is plotted for different 
external driving forces $f=5$ (green dashed line), 10 (orange solid line) and 20 (blue circles) as a function of 
the chain length, $N_0$, for fixed values of persistence length $\ell_{{p}}=25$ and pore friction $\eta_{{p}}=4$.
As can be seen, the value of $\alpha$ in the very short chain limit 
$N_0 / \tilde{\ell}_{\textrm{p}} < 1$, and for the Gaussian regime and beyond it, does not change 
if the external driving force varies from 20 to 5, while for $1<N_0 / \tilde{\ell}_{\textrm{p}} < 4$ the 
values of $\alpha$ for different values of the force $f=5$, 10 and 20 are not the same. 

\section{Scaling of the translocation time}\label{Scaling_translocation_time}

Following Ref.~\!\cite{jalal2014}, to obtain an analytical form for the translocation time we assume that only the external driving force 
$\tilde{f}$ contributes to the total force in the BD Eq.~\!(1) in the main article. This leads to reduction of 
Eq.~\!(2) in the main article to $\tilde{\phi}(\tilde{t})= \tilde{f} / \big[ \tilde{R}(\tilde{t}) + \tilde{\eta}_{\textrm{p}} + 
\tilde{\eta}_{_\textrm{TS}} \big]$. Then the total translocation time $\tilde{\tau}$ that is the sum of 
TP ($\tilde{\tau}_\mathrm{tp}$) and PP ($\tilde{\tau}_\mathrm{pp}$) times can be written as
\begin{equation}
\tilde{\tau} = \frac{1}{\tilde{f}} \bigg[ \int_0^{N_0} \tilde{R}_N dN 
+ \tilde{\eta}_{\textrm{p}} N_0 \bigg] + \tilde{\tau}_{\textrm{TS}},
\label{scaling_trans_time_SS_3}
\end{equation}
where $\tilde{\tau}_{\textrm{TS}} = \big[ \int_0^{N_0}\tilde{\eta}_{_\textrm{TS}} dN + 
\int_0^{\tilde{R}_{N_0}} (\tilde{\eta}_{_\textrm{TS,pp}} - \tilde{\eta}_{_\textrm{TS,tp}}) d\tilde{R} \big]/\tilde{f}$ is 
the contribution from the {\it trans} side friction to the total translocation time.
The second term in $\tau_{TS}$ is due to non-monotonic behavior of the trans-side friction 
$\tilde{\eta}_{\textrm{TS}}$ in the TP and PP stages, as demonstrated in Fig.~\!\ref{fig:trans-side-friction-S1}.
Here, for the TP stage the conservation of mass is $N= \tilde{s}+ \tilde{l}$ and
the TP time can be obtained by integration of $N$ from $0$ to $N_0$, while in the 
PP stage the conservation of the mass is $N= \tilde{s}+ \tilde{l}=N_0$ and
the PP time is solved by integration of $\tilde{R}$ from $\tilde{R}_{N_0}$ to zero.

In the rod-like limit the end-to-end distance of the chain is given by $\tilde{R}_N = N$. For the rod-like polymer 
the number of mobile monomers on the {\it cis} side is given by $\tilde{l} = \tilde{R}$, while on the {\it trans} side 
it is $\tilde{s}$. As the chain is stiff the TP time is much smaller than the total translocation time, i.e. 
$\tilde{\tau}_{\textrm{tp}} \ll \tilde{\tau}$, therefore the TP stage can be ignored. In the PP stage, as 
$N= \tilde{s}+ \tilde{l}=N_0$, one sets the condition $dN/d\tilde{t} = 0$ and integrates $\tilde{R}$ from 
$\tilde{R}_{N_0}$ to zero to obtain the PP time. Then, the translocation time becomes
$\tilde{\tau} = \tilde{\tau}_\mathrm{pp} = \frac{1}{\tilde{f}} \int_{0}^{\tilde{R}_{N_0}} 
d \tilde{R}  \big[ \tilde{R} + \tilde{\eta}_{\textrm{p}} + \tilde{\eta}_{_\textrm{TS}} (\tilde{t}) ]$.
Knowing $\tilde{\eta}_{\textrm{TS}} (\tilde{t}) = \tilde{s} = N_0 - \tilde{l}$ together with $\tilde{l} = \tilde{R}$ 
yield the final scaling form as
\begin{equation}
\tilde{\tau} = \frac{1}{\tilde{f}} \bigg[ \tilde{\eta}_{\textrm{p}} N_0 + N_0^2 \bigg] .
\label{scaling_trans_time_Rod_like_2}
\end{equation}
Similarly to the flexible case, the pore friction
term causes a significant correction to asymptotic scaling and the corresponding effective exponents for intermediate values
of $N_0$ will be between unity and two.